\documentclass[aps,showkeys,twocolumn,preprintnumbers,amsmath,amssymb,superscriptaddress,floatfix,nofootinbib]{revtex4}
\usepackage{graphicx,color,dcolumn,booktabs,bm}
\usepackage{amsmath}
\usepackage{graphicx}
\usepackage{longtable,lscape}
\usepackage{txfonts}
\usepackage{overpic}
\usepackage{epsfig}
\usepackage{amssymb}
\usepackage{rotating}
\usepackage{epstopdf}
\usepackage{appendix}
\usepackage{indentfirst}
\usepackage{feynmf}   
\usepackage{slashed}  
\usepackage{cases}
\usepackage{color}
\usepackage{multirow}
\usepackage{graphicx,color,dcolumn,booktabs,bm}
\usepackage{cases}
\usepackage{array}

\graphicspath{{Figures/}} %

\usepackage[colorlinks, citecolor=blue,anchorcolor=red,menucolor=red, linkcolor=red,filecolor=red,runcolor=red,urlcolor=blue,frenchlinks=red]{hyperref}

\begin{document}
\title{Exploring bottom-charmed molecular tetraquarks with the complex scaling method}
\author{Qing-Fu Song}\email{242201003@csu.edu.cn}

\affiliation{School of Physics, Central South University, Changsha 410083, China}

\author{Wei Liang}

\affiliation{School of Physics, Central South University, Changsha 410083, China}

\author{Qi-Fang L\"{u}}\email{lvqifang@hunnu.edu.cn}

\affiliation{Department of Physics, Hunan Normal University, Changsha 410081, China}
\affiliation{Key Laboratory of Low-Dimensional Quantum Structures and Quantum Control of Ministry of Education, Changsha 410081, China}
\affiliation{Key Laboratory for Matter Microstructure and Function of Hunan Province, Hunan Normal University, Changsha 410081, China}

\author{Xiao-Nu Xiong}\email{xnxiong@csu.edu.cn}

\affiliation{School of Physics, Central South University, Changsha 410083, China}

\begin{abstract}
In this work, we adopt the one-boson exchange model to analyze the $D^{(*)}_{(s)}B^{(*)}_{(s)}$ interactions from a coupled-channel perspective. For the $I(J^{P})=1(1^{+})$ $DB^{*}/D^*B/D^*B^{*}$ system, employing the complex scaling method, we obtain a loosely bound state and two resonant states located below the $DB^{*}$, $D^*B$, and $D^*B^{*}$ thresholds, respectively. Assuming that the charmonium-like states $T_{c\bar{c}1}(3900)$ and $T_{c\bar{c}1}(4020)$, as well as the bottomonium-like states $T_{b\bar{b}1}(10610)$ and $T_{b\bar{b}1}(10650)$, are molecular states near their respective thresholds, we suggest that the predicted states in the $I(J^P)=1(1^+)$ $D^{(*)}_{(s)}B^{(*)}_{(s)}$ system could be their bottom-charmed analogues. In addition, we extend our analysis to other $D^{(*)}_{(s)}B^{(*)}_{(s)}$ combinations with various quantum numbers. Our results suggest that some exotic structures may exist in the bottom-charmed sector. Finally, we collect the possible decay channels for future experimental observations. The $B_c/B_c^*$ plus a light meson are the ideal final state to search for the bound states, while the $D_{(s)}^{(*)}B_{(s)}^{(*)}$ channels are suitable for the resonances. We highly recommend that the future experiments hunt for these bottom-charmed exotic particles.

\end{abstract}


\maketitle
\section{introduction}
 The discovery of hidden charm state $X(3872)$ with $J^{PC}=1^{++}$ in the process $B^{\pm}$$\to $$K^{\pm}\pi^{+}\pi^{-}J/\psi$ by the Belle Collaboration in 2003 caused a series of interesting consequences for the establishment of hadronic spectra~\cite{Belle:2003nnu}. Since $X(3872)$ lies very close to the $D \bar{D}^{*} + h.c.$ threshold, it was naturally interpreted as a $D \bar D^{*}$ hadronic molecular state. This discovery sparks further interest in exotic hadrons, especially in the charm sector. In 2013, the BESIII~\cite{BESIII:2013ris} and Belle~\cite{Belle:2013yex} Collaborations observed a charged charmonium-like state named $T_{c\bar{c}1}(3900)$, which generates even more attention. As $T_{c\bar{c}1}(3900)$ is located slightly above the $D \bar D^{*} +h.c.$ threshold, it can be identified as the isospin partner of $X(3872)$. According to heavy quark spin symmetry, there should exist charmonium-like states in $D^*\bar{D}^{*}$ channel. Indeed, an isovector state $T_{c\bar{c}1}(4020)$ and an isoscalar state $X(4013)$ were subsequently discovered by the BESIII~\cite{BESIII:2013ouc} and Belle~\cite{Belle:2021nuv} Collaborations, respectively. Meanwhile, for the bottom sector, two bottomonium-like states $T_{c\bar{c}1}(10610)$ and $T_{c\bar{c}1}(10650)$ were reported by the Belle Collaboration in 2011~\cite{Belle:2011aa}. These states with $I(J^{PC})=1(1^{+-})$ have masses of $10607.2^{+2}_{-2}$ and $10652.2^{+1.5}_{-1.5}$$\rm~MeV$, and widths of $18.4^{+2.4}_{-2.4}$ and $11.5^{+2.2}_{-2.2}$$\rm~MeV$, respectively. These findings provide additional evidence for the existence of exotic hadrons in both the charm and bottom sectors. Meanwhile, a series of new phenomenological studies on the $XYZ$ and $P_c/T_{cc}$ states have been reported in references~\cite{Liu:2019zoy, Wang:2025sic,Chen:2016qju,Liu:2024uxn,Hosaka:2016pey,Lebed:2016hpi,Ali:2017jda,
Esposito:2016noz,Chen:2022asf,Dong:2017gaw,Guo:2017jvc,Olsen:2017bmm,
Karliner:2017qhf,Brambilla:2019esw,Bicudo:2022cqi,Mai:2022eur,
Barabanov:2020jvn}.

The $B_c$ meson was first observed in 1998 by the CDF Collaboration through the decay modes $B_c^{\pm} \to J/\psi \ell^{\pm} X$ and $B_c^{\pm} \to J/\psi \ell^{\pm} \bar{\nu}_{\ell}$ in $p\bar{p}$ collisions. The mass was measured to be $6.40 \pm 0.39 \pm 0.13 \rm{GeV}$~\cite{CDF:1998axz}, and this result was later confirmed by the LHCb and D0 Collaborations~\cite{LHCb:2012ihf,D0:2008bqs}. According to the Particle Data Group (PDG)~\cite{ParticleDataGroup:2024cfk}, the $B_c$ meson has a mass of $m = 6274.47\pm0.32~\rm MeV$ and quantum numbers $I(J^P)=0(0^-)$. In 2019, the CMS Collaboration reported the discovery of radially excited states $B^{+}_c(2^1S_0)$ and $B_c^{*+}(2^3S_1)$ in the $B_c^+ \pi^+ \pi^-$ invariant mass spectrum. These two states were found to have a mass difference of $m(2^1S_0)-m(2^3S_1)=29\pm1.5\pm0.7~\rm MeV$~\cite{CMS:2019uhm}. Also in 2019, the LHCb Collaboration observed two excited $c\bar{b}$ states in the same decay channel. The masses of these two states were measured to be $6841.2 \pm 0.6 \pm 0.1 \pm 0.8 ~\rm{MeV}$ and $6872.1 \pm 1.3 \pm 0.1 \pm 0.8 ~\rm{MeV}$, respectively~\cite{LHCb:2019bem}. Recently, the LHCb Collaboration reports the first observation of orbitally excited \( B_c \) mesons, with two narrow peaks observed at \( 6704.8 \pm 5.5 \pm 2.8 \pm 0.3 \,\mathrm{MeV}/c^2 \) and \( 6752.4 \pm 9.5 \pm 3.1 \pm 0.3 \,\mathrm{MeV}/c^2 \) in the \( B_c^+ \gamma \) mass spectrum, based on 9~fb\(^{-1} \) of \( pp \) collision data. The structure's significance exceeds $7\sigma$, and the results are consistent with theoretical predictions for the lowest \( P \)-wave \( B_c^+ \) excitations~\cite{LHCb:2025uce,LHCb:2025ubr}.
 However, until now, there are no more significant experimental signals for hadrons containing both charm and bottom quarks, which include tetraquarks.  

The spectroscopy of traditional bottom-charmed mesons has been extensively discussed in the framework of quark models~\cite{Godfrey:1985xj,Kwong:1990am,Eichten:1994gt,Zeng:1994vj,Gupta:1995ps,Fulcher:1998ka,Ikhdair:2003ry,Ebert:2002pp,Godfrey:2004ya,Ikhdair:2004hg,Ikhdair:2004tj,Soni:2017wvy,Eichten:2019gig,Li:2019tbn}, lattice QCD~\cite{Allison:2004be,Dowdall:2012ab,Mathur:2018epb}, and QCD sum rules~\cite{Wang:2012kw,Gershtein:1994dxw}. The bottom-charmed molecular tetraquarks, known as $B_c$-like states, have not been observed experimentally. However, the first principles of quantum chromodynamics (QCD) do not prohibit their existence. Theoretically, many works focus on their existence and properties. In Ref.~\cite{Wu:2018xdi}, the author calculated the mass spectra of the $cq\bar{b}\bar{q}$, $cs\bar{b}\bar{q}$, and $cs\bar{b}\bar{s}$ components using the chromomagnetic interaction model. Many references have extensively investigated the properties of $B_c$-like states within the QCD sum rules method and QCD light-cone sum rules~\cite{Zhang:2009em,Albuquerque:2012rq,Ozdem:2024rrg,Wang:2020jgb,Agaev:2017uky,Ozdem:2022eds,Chen:2013aba,Agaev:2016dsg,Zhang:2009vs}. For instance, in Ref.~\cite{Zhang:2009vs}, the authors examined the masses of ${Q\bar{q}}{\bar{Q}^{(\prime)}q}$ molecules and obtained some bound states within the QCD sum rules framework for the bottom-charmed sector. The chiral unitary approach and hidden gauge symmetry Lagrangians have also been used to study $B_c$-like molecular states~\cite{Wang:2023jeu,Sakai:2017avl}. Additionally, in Ref.~\cite{Sun:2012sy}, the interaction between $B_{(s)}^{(*)}$ and $D_{(s)}^{(*)}$ mesons was studied using the one-boson exchange (OBE) model. Their results suggested the possible existence of molecular states. In Ref.~\cite{Wang:2023bek}, the authors also studied radiative decays and magnetic moments for bottom-charmed molecular tetraquarks.

\begin{figure}
    \centering
    \includegraphics[scale=0.5]{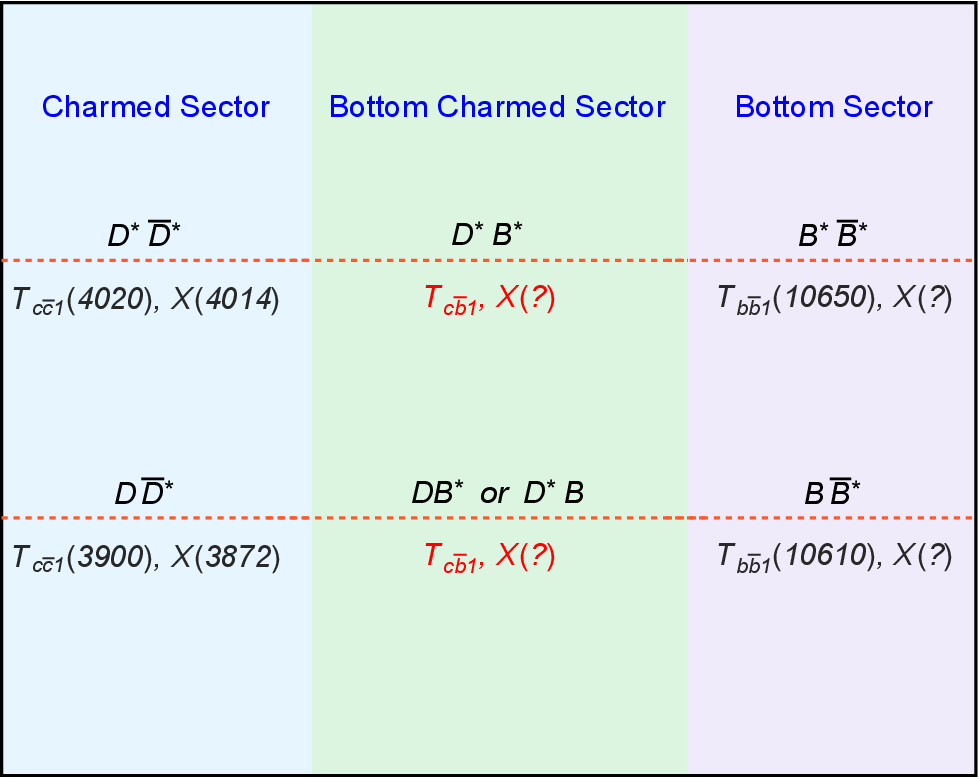}
    \caption{Comparison of $J^{P}=1^{+}$ states for charmed, bottom-charmed, and bottom sectors.}
    \label{spectrum}
\end{figure}

However, we are still far from establishing the complete spectroscopy for $B_c$-like molecular tetraquark states, and our understanding of them remains scarce. The exploration of bottom-charmed molecular tetraquark states may provide new insights into the nature of the exotic states observed near the $D^{(*)}\bar{D}^{*}$ and $B^{(*)}\bar{B}^{*}$ thresholds. On the other hand, heavy quark symmetry (HQS) implies that the strong interaction dynamics for heavy quarks are similar, regardless of whether the quark is charm or bottom. As illustrated in Figure~\ref{spectrum}, HQS allows for the existence of exotic hadronic states in the bottom-charmed sector, expected to be counterparts to those already observed in the charm and bottom sectors. Furthermore, coupled channel effects and $S-D$ wave mixing effects are expected to play an important role in the production of these molecular states. Thus, it is urgent to study the $D^{(*)}_{(s)}B^{(*)}_{(s)}$ systems systematically using the coupled channel approach and explore possible excited resonances along with bound states.

Recently, we have systematically explored the properties of the hidden bottom molecular tetraquarks and the bottom-strange molecular pentaquarks by adopting the one-boson exchange model~\cite{Song:2024ngu,Song:2025yut}. We utilize the Gaussian expansion method~\cite{Hiyama:2003cu,Hiyama:2018ivm} and the complex scaling method~\cite{Moiseyev:1998gjp, Ho:1983lwa} to solve the coupled-channel Schr\"odinger equation and obtain poles. In the present work, we systematically investigate bottom-charmed molecular tetraquark states using the same procedure. First, we investigate $DB^*/D^*B/D^{*}B^{*}$ with $I(J^{P})=1(1^{+})$, which helps determine the free parameter cutoff $\Lambda$. According to our estimations, we predict a bound state below the $DB^*$ threshold and two resonances below the $D^*B$ and $D^*B^*$ thresholds, respectively. Then, we extend our explorations to various other $D^{(*)}_{(s)}B^{(*)}_{(s)}$ systems with different quantum numbers. In summary, we predict three molecular states for the $I(J^{P})=0(0^+)$ system, two for the $I(J^P)=0(1^+)$ system, one for the $I(J^P)=0(2^+)$ system, one for the $I(J^P)=1(0^+)$ system, and three for the $I(J^P)=1(1^+)$ system. We hope that our predictions can offer useful information for future experimental observations.

This paper is organized as follows. The formalism of effective interactions and complex scaling method are briefly introduced in Sec.~\ref{model}. We present the numerical results and discussions for the $D^{(*)}_{(s)}B^{(*)}_{(s)}$ systems in Sec.~\ref{results}. A summary is given in the last section.

\section{Formalism}\label{model}
\subsection{The effective interactions}
In the present work, we employ the OBE model to obtain the effective potential of the $D^{(*)}_{(s)}B^{(*)}_{(s)}$ systems. The OBE model describes meson-meson interaction by exchanging the light scalar, pseudoscalar and vector mesons, which is successfully applied to explain the dynamic mechanisms of molecular states.  
Under the heavy quark symmetry and chiral symmetry, the relevant effective Lagrangians can be expressed as~\cite{Yan:1992gz,Wise:1992hn,Burdman:1992gh,Casalbuoni:1996pg}
\begin{eqnarray}
{\mathcal L}&=&g_{\sigma}\left\langle H^{(Q)}_a\sigma\overline{H}^{(Q)}_a\right\rangle+g_{\sigma}\left\langle \overline{H}^{(\bar{Q})}_a\sigma H^{(\bar{Q})}_a\right\rangle\nonumber\\
&&+ig\left\langle H^{(Q)}_b{\mathcal A}\!\!\!\slash_{ba}\gamma_5\overline{H}^{\,({Q})}_a\right\rangle+ig\left\langle \overline{H}^{(\bar{Q})}_a{\mathcal A}\!\!\!\slash_{ab}\gamma_5 H^{\,(\bar{Q})}_b\right\rangle\nonumber\\
&&+\left\langle iH^{(Q)}_b\left(\beta v^{\mu}({\mathcal V}_{\mu}-\rho_{\mu})+\lambda \sigma^{\mu\nu}F_{\mu\nu}(\rho)\right)_{ba}\overline{H}^{\,(Q)}_a\right\rangle\nonumber\\
&&-\left\langle i\overline{H}^{(\bar{Q})}_a\left(\beta v^{\mu}({\mathcal V}_{\mu}-\rho_{\mu})-\lambda \sigma^{\mu\nu}F_{\mu\nu}(\rho)\right)_{ab}H^{\,(\bar{Q})}_b\right\rangle,
\end{eqnarray}
where $a$ and $b$ is the flavor indices, and $v^{\mu}=(1, \bf{0})$ is the four-velocity. The vector current is 
\begin{equation}
       \mathcal{V}_{\mu} = \frac{1}{2}(\xi^{\dag}\partial_{\mu}\xi+\xi\partial_{\mu}\xi^{\dag}),
\end{equation}
the axial current is 
\begin{equation}
    A_{\mu}=\frac{1}{2}(\xi^{\dag}\partial_{\mu}\xi-\xi\partial_{\mu}\xi^{\dag}),
\end{equation}
and strength tensor of vector field is
\begin{equation}
F_{\mu\nu}(\rho)=\partial_{\mu}\rho_{\nu}-\partial_{\nu}\rho_{\mu}+[\rho_{\mu},\rho_{\nu}],
  \end{equation}
where $\xi=\text{exp}(i{\mathbb{P}}/f_{\pi})$ and $\rho_{\mu}=ig_{V}{\mathcal{V}}_{\mu}/\sqrt{2}$. The $f_{\pi}=132$$\rm~MeV$ is the pion decay constant, and then $g_V=m_{\rho}/f_{\pi}=5.8$~\cite{
Bando:1987br}. The $\mathbb{P}$ and $\mathbb{V}$ stand for the matrices of light pseudoscalar and vector mesons, respectively,  
\begin{eqnarray}
\left.\begin{array}{c} {\mathbb{P}} = {\left(\begin{array}{ccc}
       \frac{\pi^0}{\sqrt{2}}+\frac{\eta}{\sqrt{6}} &\pi^+ &K^+\\
       \pi^-       &-\frac{\pi^0}{\sqrt{2}}+\frac{\eta}{\sqrt{6}} &K^0\\
       K^-         &\bar K^0   &-\sqrt{\frac{2}{3}} \eta     \end{array}\right)},\\
{\mathbb{V}} = {\left(\begin{array}{ccc}
       \frac{\rho^0}{\sqrt{2}}+\frac{\omega}{\sqrt{2}} &\rho^+ &K^{*+}\\
       \rho^-       &-\frac{\rho^0}{\sqrt{2}}+\frac{\omega}{\sqrt{2}} &K^{*0}\\
       K^{*-}         &\bar K^{*0}   & \phi     \end{array}\right)}.
\end{array}\right.
\end{eqnarray}
The $H^{(Q)}_a$, $H^{(\bar Q)}_a$, $\bar H^{(Q)}_a$, and $\bar H^{(\bar Q)}_a$ represent the fields of heavy-light mesons   and can be written as 
\begin{eqnarray}
H^{(Q)}_a&=&(1+{v}\!\!\!\slash)(\mathcal{P}^{*(Q)\mu}_a\gamma_{\mu}-\mathcal{P}^{(Q)}_a\gamma_5)/2,\\
H^{(\overline{Q})}_a &=& (\bar{\mathcal{P}}^{*(\overline{Q})\mu}_a\gamma_{\mu}-\bar{\mathcal{P}}^{(\overline{Q})}_a\gamma_5)
(1-{v}\!\!\!\slash)/{2},\\
\overline{H}&=&\gamma_0H^{\dagger}\gamma_0.
\end{eqnarray}
More explicitly, the effective Lagrangian depicting the couplings of light mesons and heavy-light mesons can be written as~\cite{Chen:2022dad,Li:2012cs,Sun:2011uh}
\begin{eqnarray}
	\mathcal{L}_{\mathcal{P} \mathcal{P} \mathbb{V}}&= & -\sqrt{2} \beta g_{V} \mathcal{P}_{b} \mathcal{P}_{a}^{\dagger} v \cdot \mathbb{V}_{b a}+\sqrt{2} \beta g_{V} \widetilde{\mathcal{P}}_{a}^{\dagger} \widetilde{\mathcal{P}}_{b} v \cdot \mathbb{V}_{a b},\\
	\mathcal{L}_{\mathcal{P}^{*} \mathcal{P} \mathbb{V}}&= & -2 \sqrt{2} \lambda g_{V} v^{\lambda} \varepsilon_{\lambda \mu \alpha \beta}\left(\mathcal{P}_{b} \mathcal{P}_{a}^{* \mu \dagger}+\mathcal{P}_{b}^{* \mu} \mathcal{P}_{a}^{\dagger}\right)\left(\partial^{\alpha} \mathbb{V}^{\beta}\right)_{b a} \nonumber\\
	& -&2 \sqrt{2} \lambda g_{V} v^{\lambda} \varepsilon_{\lambda \mu \alpha \beta}\left(\widetilde{\mathcal{P}}_{a}^{* \mu \dagger} \widetilde{\mathcal{P}}_{b}+\widetilde{\mathcal{P}}_{a}^{\dagger} \widetilde{\mathcal{P}}_{b}^{* \mu}\right)\left(\partial^{\alpha} \mathbb{V}^{\beta}\right)_{a b},\\
	\mathcal{L}_{\mathcal{P}^{*} \mathcal{P}^{*} \mathbb{V}}&= & \sqrt{2} \beta g_{V} \mathcal{P}_{b}^{*} \cdot \mathcal{P}_{a}^{* \dagger} v \cdot \mathbb{V}_{b a} \nonumber\\
	& -&i 2 \sqrt{2} \lambda g_{V} \mathcal{P}_{b}^{* \mu} \mathcal{P}_{a}^{* v^{\dagger}}\left(\partial_{\mu} \mathbb{V}_{v}-\partial_{v} \mathbb{V}_{\mu}\right)_{b a} \nonumber\\
	& -&\sqrt{2} \beta g_{V} \widetilde{\mathcal{P}}_{a}^{* \dagger} \widetilde{\mathcal{P}}_{b}^{*} v \cdot \mathbb{V}_{a b} \nonumber\\
	& -&i 2 \sqrt{2} \lambda g_{V} \widetilde{\mathcal{P}}_{a}^{* \mu \dagger} \widetilde{\mathcal{P}}_{b}^{* v}\left(\partial_{\mu} \mathbb{V}_{v}-\partial_{v} \mathbb{V}_{\mu}\right)_{a b},\\
    \mathcal{L}_{\mathcal{P}^{*} \mathcal{P}^{*} \mathbb{P}}&= & -i \frac{2 g}{f_{\pi}} v^{\beta} \varepsilon_{\beta \mu \alpha v} \mathcal{P}_{b}^{* \mu} \mathcal{P}_{a}^{* \nu \dagger} \partial^{\alpha} \mathbb{P}_{b a}\nonumber \\
& +&i \frac{2 g}{f_{\pi}} v^{\beta} \varepsilon_{\beta \mu \alpha \nu} \widetilde{\mathcal{P}}_{a}^{* \mu \dagger} \widetilde{\mathcal{P}}_{b}^{* v} \partial^{\alpha} \mathbb{P}_{a b},\\
	\mathcal{L}_{ \mathcal{P} \mathcal{P{\sigma}}} & =&-2 g_{s} \mathcal{P}_{b} \mathcal{P}_{b}^{\dagger} \sigma-2 g_{s} \widetilde{\mathcal{P}}_{b} \widetilde{\mathcal{P}}_{b}^{\dagger} \sigma, \\
	\mathcal{L}_{\mathcal{P}^{*} \mathcal{P}^{*} \sigma} & =&2 g_{s} \mathcal{P}_{b}^{*} \cdot \mathcal{P}_{b}^{* \dagger} \sigma+2 g_{s} \widetilde{\mathcal{P}}_{b}^{*} \cdot \widetilde{\mathcal{P}}_{b}^{* \dagger} \sigma,
\end{eqnarray}
where the relevant parameters and details can be found in Refs. ~\cite{Isola:2003fh, Chen:2022dad}.

Using the  given  Lagrangians, we can deduce the  relevant interaction potentials of these investigated systems in the momentum space straightforwardly. Based on the Breit approximation, the effective potential reads
\begin{eqnarray}\label{breit}
\mathcal{V}^{M_{1}M_{2}\to M_{3}M_{4}}(\bm{q}) &=&
          -\frac{\mathcal{M}(M_{1}M_{2}\to M_{3}M_{4})}
          {4\sqrt{m_{1}m_{2}m_{3}m_{4}}},
\end{eqnarray}
where $\mathcal{M}(M_{1}M_{2}\to M_{3}M_{4})$ denotes the scattering amplitude for the $\mathcal{M}(M_{1}M_{2}\to M_{3}M_{4})$ process, and $m_{i}$ is the mass of the meson $M_i$.  
One can obtain the final effective potential in position space after performing the Fourier transformation
\begin{eqnarray}
\mathcal{V}(\bm{r}) &=&
          \int\frac{d^3\bm{q}}{(2\pi)^3}e^{i\bm{q}\cdot\bm{r}}
          \mathcal{V}(\bm{q})\mathcal{F}^2(q^2,m_i^2),\\
    \mathcal{F}(q^2,m_i^2)&=& (\Lambda^2-m_i^2)/(\Lambda^2-q^2)\label{FF},
\end{eqnarray}
since the hadrons are not point particles, the form factor with the cutoff parameter $\Lambda$ is introduced to account for the inner structures of them.
Then, we can obtain the flavor-independent sub-potentials
\begin{eqnarray}
V^{a}_{v}&=& -\frac{1}{2}\beta^{2}g_{V}^{2}Y(\Lambda,m_{v},r)\nonumber,\\
V^{a}_{\sigma}&=&-g_{s}^{2}Y(\Lambda,m_{\sigma},r),
\end{eqnarray}
for $PP\to PP$ process,
\begin{eqnarray}
V^{c}_{p}&=&-\frac{1}{3}\frac{g^{2}}{f_{\pi}^{2}} \mathcal{Z}_{\Lambda, m_{p}},\nonumber\\
V^{c}_{v}&=&\frac{2}{3}\lambda^{2}g_{V}^{2}\mathcal{Z}_{\Lambda, m_{v}}^{\prime},
\end{eqnarray}
for $PP\to P^{*}P^{*}$ process,
\begin{eqnarray}
    V^{d}_{\sigma}&=&-g_{s}^{2}\mathcal{Y}_{\Lambda, m_{\sigma}},\nonumber\\
    V^{d}_{v}&=&-\frac{1}{2}\beta^{2}g_{V}^{2}\mathcal{Y}_{\Lambda, m_{v}},
\end{eqnarray}
for $ P^{}P^{*}\to P^{}P^{*}$ process, 
    \begin{eqnarray}
V^{e}_{p}&=&-\frac{1}{3}\frac{g^{2}}{f_{\pi}^{2}} \mathcal{Z}_{\Lambda, m_{p}},\nonumber\\
V^{e}_{v}&=&\frac{2}{3}\lambda^{2}g_{V}^{2}\mathcal{Z}_{\Lambda, m_{v}}^{\prime},
\end{eqnarray}
for $ P^{}P^{*}\to P^{*}P^{}$ process, and
\begin{eqnarray}\nonumber
V_{p}^{f} &=& -\frac{1}{3}\frac{g^2}{f_{\pi}^2}\mathcal{Z}_{\Lambda_{i},m_{i}}^{ij},\nonumber\\
    V_{v}^{f} &=&\frac{2}{3}\lambda^2g_V^2
    \mathcal{X}_{\Lambda_{i},m_{i}}^{ij},\nonumber\\
\end{eqnarray}
for $PP^{*}\to P^{*}P^{*}$ process. 
\begin{eqnarray}
       V_{p}^{g} &=& -\frac{1}{3}\frac{g^2}{f_{\pi}^2}\mathcal{Z}_{\Lambda_{i},m_{i}}^{ji},\nonumber\\
    V_{v}^{g} &=&\frac{2}{3}\lambda^2g_V^2\mathcal{X}_{\Lambda_{i},m_{i}}^{ji},
\end{eqnarray}
for $P^{*}P\to P^{*}P^{*}$ process, \begin{eqnarray}
V^{h}_{p}&=&-\frac{1}{3}\frac{g^{2}}{f_{\pi}^{2}} \mathcal{Z}_{\Lambda, m_{p}},\nonumber\\
V^{h}_{\sigma}&=&-g_{s}^{2}\mathcal{Y}_{\Lambda, m_{\sigma}},\nonumber\\
V^{h}_{v}&=&-\frac{1}{2}\beta^{2}g_{V}^{2}\mathcal{Y}_{\Lambda, m_{v}}+\frac{2}{3}\lambda^{2}g_{V}^{2}\mathcal{Z}_{\Lambda, m_{v}}^{\prime},
\end{eqnarray}
for $ P^{*}P^{*}\to P^{*}P^{*}$ process. 
Some explicit formulas for the potentials can be expressed as
\begin{eqnarray}
\mathcal{Z}_{\Lambda,m_a}^{ij}&=&\Bigg(\mathcal{E}_{ij}\nabla^{2}+\mathcal{F}_{ij}r\frac{\partial}{\partial r}\frac{1}{r}\frac{\partial}{\partial r}\Bigg) Y(\Lambda,m,r),\nonumber\\\
\mathcal{Z}_{\Lambda,m_a}^{\prime ij}&=&\Bigg(2\mathcal{E}_{ij}\nabla^{2}-\mathcal{F}_{ij}r\frac{\partial}{\partial r}\frac{1}{r}\frac{\partial}{\partial r}\Bigg) Y(\Lambda,m,r),\nonumber\\\
\mathcal{X}^{ij}_{\Lambda,m_a}&=&\Bigg(-2\mathcal{E}_{ij}\nabla^{2}-(\mathcal{F}^{\prime}_{ij}-\mathcal{F}^{\prime\prime}_{ij})r\frac{\partial}{\partial r}\frac{1}{r}\frac{\partial}{\partial r}\Bigg) Y(\Lambda,m,r),\nonumber\\\
\mathcal{Y}^{ij}_{\Lambda,m_a}&=&\mathcal{D}_{ij} Y(\Lambda,m,r),\nonumber\\\
Y(\Lambda,m,r)&=&\frac{1}{4\pi r}(e^{-mr}-e^{-\Lambda r})-\frac{\Lambda^2-m^2}{8\pi\Lambda}e^{-\Lambda r}.
\end{eqnarray}
Here, $\mathcal{D}_{ij}$, $\mathcal{E}_{ij}$, and $\mathcal{F}_{ij}$ represent  the operators for spin-spin coupling and  the tensor forces and relate with polarization vector  $\epsilon_i$ and Pauli matrix $\sigma$, respectively. 
Finally, the total effective potentials for $D^{(*)}_{(s)}B^{(*)}_{(s)}$ systems can be expressed by the combinations of these sub-potentials~\cite{Chen:2022dad,Li:2012cs,Sun:2011uh} and summarized in Table~\ref{total}. Also, the relevant parameters are collected in Table~\ref{parameters}~\cite{Bando:1987br,Wang:2021yld,Chen:2022dad,Wang:2019aoc,Machleidt:1987hj}.

\begin{table*}[!htbp]
	\renewcommand\arraystretch{1.4}
	\caption{\label{total} The effective  potentials for $D^{(*)}_{(s)}B^{(*)}_{(s)}$ systems.}
	\begin{ruledtabular}\footnotesize
		\begin{tabular}{cccccccccc}
			&$I(J^{PC})=0(0^{+})$&$DB$
			&$D_sB_s$
			&$D^*B^*$
			&$D_s^*B_s^*$&&
			\\
			&$DB$&$V^{a}_{\sigma}+\frac{3}{2}V_{\rho}^{a}+\frac{1}{2}V_{\omega}^{a}$&$\sqrt{2}V_{K^{*}}^{a}$& $\frac{1}{6}V^{b}_{\eta}+\frac{3}{2}V^{b}_{\pi}+\frac{3}{2}V_{\rho}^{b}+\frac{1}{2}V_{\omega}^{b}$&$\sqrt{2}V_{K^{*}}^{b}+\sqrt{2}V_{K}^{b}$&&\\
			&$D_sB_s$&&$V^{a}_{\sigma}+V^{a}_{\phi}$&$\sqrt{2}V_{K^{*}}^{b}+\sqrt{2}V_{K}^{b}$&$\frac{2}{3}V_{\eta}^{b}+V_{\phi}^{b}$&&\\
			&$D^*B^*$ &&&$V_{\sigma}^{c}+\frac{3}{2}V_{\pi}^{c}+\frac{1}{6}V_{\eta}^{c}+\frac{3}{2}V_{\rho}^{c}+\frac{1}{2}V_{\omega}^{c}$&$\sqrt{2}V_{K^{*}}^{c}+\sqrt{2}V_{K}^{c}$\\
			&$D_s^*B_s^*$ &&&&$V_{\sigma}^{c}+\frac{2}{3}V_{\eta}^{c}+V_{\phi}^{c}$ \\
			\hline
			
			&$I(J^{PC})=0(1^{+})$&$DB^*$&$D^{*}B$
			&$D^*B^*$
   &$D_sB_s^{*}$
			&$D_s^{*}B_s$		
			&$D_s^*B_s^*$
			\\
			&\multirow{1}{*}{$DB^{*}$}&$V^{d}_{\sigma}+\frac{3}{2}V_{\rho}^{d}+\frac{1}{2}V_{\omega}^{d}$&$\frac{3}{2}V_{\pi}^{f}+\frac{1}{6}V_{\eta}^{f}+\frac{3}{2}V_{\rho}^{f}+\frac{1}{2}V_{\omega}^{f}$&$V_{\sigma}^{c}+\frac{3}{2}V_{\pi}^{c}+\frac{1}{6}V_{\eta}^{c}+\frac{3}{2}V_{\rho}^{c}+\frac{1}{2}V_{\omega}^{c}$&$\sqrt{2}V_{K^{*}}^{d}$&$\sqrt{2}V_{K^{*}}^{d}$&$\sqrt{2}V_{K^{*}}^{f}+\sqrt{2}V_{K}^{f}$\\
			&\multirow{1}{*}{$D^{*}B$}&$\frac{3}{2}V^{e}_{\pi}+\frac{1}{6}V_{\eta}^{e}+\frac{3}{2}V^{e}_{\rho}+\frac{1}{2}V^{e}_{\omega}$&$\frac{3}{2}V_{\pi}^{g}+\frac{1}{6}V_{\eta}^{g}+\frac{3}{2}V_{\rho}^{g}+\frac{1}{2}V_{\omega}^{g}$&$\sqrt{2}V_{K^{*}}^{e}+\sqrt{2}V_{K}^{e}$&$\sqrt{2}V_{K^{*}}^{g}+\sqrt{2}V_{K}^{g}$\\
			&\multirow{1}{*}{$D^{*}B^{*}$}&&\multirow{1}{*}{$V_{\sigma}^{c}+\frac{3}{2}V_{\pi}^{c}+\frac{1}{6}V_{\eta}^{c}+\frac{3}{2}V_{\rho}^{c}+\frac{1}{2}V_{\omega}^{c}$}&$\sqrt{2}V_{K^{*}}^{f}+\sqrt{2}V_{K}^{f}$&\multirow{1}{*}{$\sqrt{2}V_{K^{*}}^{c}+\sqrt{2}V_{K}^{c}$}\\
			&\multirow{1}{*}{$D_sB_s^*$ }&&&\multirow{1}{*}{$V^{d}_\sigma+V_{\phi}^d+\frac{2}{3}V_{\eta}^{e}+V_{\phi}$}&$\frac{2}{3}V^{f}_{\eta}+V^{f}_{\phi}$\\
		&\multirow{1}{*}{$D_s^{*}B_s$ }&&&&$\frac{2}{3}V^{g}_{\eta}+V^{g}_{\phi}$\\
			&$D_s^*B_s^*$ &&&&$V_{\sigma}^{c}+\frac{2}{3}V_{\eta}^{c}+V_{\phi}^{c}$ \\\hline
			
			&$I(J^{PC})=0(2^{+})$
			&$D^*B^*$
			&$D_s^*B_s^*$\\
			&$D^*B^*$& $V_{\sigma}^{c}+\frac{3}{2}V_{\pi}^{c}+\frac{1}{6}V_{\eta}^{c}+\frac{3}{2}V_{\rho}^{c}+\frac{1}{2}V_{\omega}^{c}$&$\sqrt{2}V_{K^{*}}^{c}+\sqrt{2}V_{K}^{c}$\\
			&$D_s^*B_s^*$& &$V_{\sigma}^{c}+\frac{2}{3}V_{\eta}^{c}+V_{\phi}^{c}$ \\
			\hline
			
			&$I(J^{PC})=1(0^{+})$&$DB$
			&$D^*B^*$
			\\
			&$DB$&$V^{d}_{\sigma}-\frac{1}{2}V_{\rho}^{d}+\frac{1}{2}V_{\omega}^{d}$& $\frac{1}{6}V^{b}_{\eta}-\frac{1}{2}-\frac{1}{2}V_{\rho}^{b}+\frac{1}{2}V_{\omega}^{b}$\\
			&$D^*B^*$ &&$V_{\sigma}^{c}-\frac{1}{2}V_{\pi}^{c}+\frac{1}{6}V_{\eta}^{c}-\frac{1}{2}V_{\rho}^{c}+\frac{1}{2}V_{\omega}^{c}$\\\hline
			
				&$I(J^{PC})=1(1^{+})$&$DB^*$&$D^{*}B$
			&$D^*B^*$\\
   &\multirow{1}{*}{$DB^*$}&$V^{d}_{\sigma}-\frac{1}{2}V_{\rho}^{d}+\frac{1}{2}V_{\omega}^{d}$&$\frac{3}{2}V_{\pi}^{f}+\frac{1}{6}V_{\eta}^{f}+\frac{3}{2}V_{\rho}^{f}+\frac{1}{2}V_{\omega}^{f}$&$-\frac{1}{2}V_{\pi}^{f}+\frac{1}{6}V_{\eta}^{f}-\frac{1}{2}V_{\rho}^{f}+\frac{1}{2}V_{\omega}^{f}$\\
			&\multirow{1}{*}{$D^{*}B$}&&$V^{d}_{\sigma}-\frac{1}{2}V_{\rho}^{d}+\frac{1}{2}V_{\omega}^{d}$&$-\frac{1}{2}V_{\pi}^{f}+\frac{1}{6}V_{\eta}^{f}-\frac{1}{2}V_{\rho}^{f}+\frac{1}{2}V_{\omega}^{f}$\\
			&$D^{*}B^{*}$&&&$V_{\sigma}^{c}-\frac{1}{2}V_{\pi}^{c}+\frac{1}{6}V_{\eta}^{c}-\frac{1}{2}V_{\rho}^{c}+\frac{1}{2}V_{\omega}^{c}$\\\hline

			&$I(J^{PC})=1(2^{+})$
			&$D^*B^*$
			\\
			&$D^*B^*$& $V_{\sigma}^{c}-\frac{1}{2}V_{\pi}^{c}+\frac{1}{6}V_{\eta}^{c}-\frac{1}{2}V_{\rho}^{c}+\frac{1}{2}V_{\omega}^{c}$\\
		\end{tabular}
	\end{ruledtabular}
\end{table*}
\begin{table}[htbp]
\caption{\label{parameters} The relevant parameters adopted in this work.}
\begin{ruledtabular}
\begin{tabular}{ccccccccc}
&Parameters
&Value
\\\hline
&$g_{V}$&5.8\\
&$g$&0.59\\
&$g_s$&2.82\\
&$\beta$&0.9\\
&$f_{\pi}$&0.132$(\textrm{GeV})$\\
&$\lambda$&0.56$(\textrm{GeV}^{-1})$\\
\end{tabular}
\end{ruledtabular}
\end{table}

\begin{figure}
    \centering
    \includegraphics[scale=0.65]{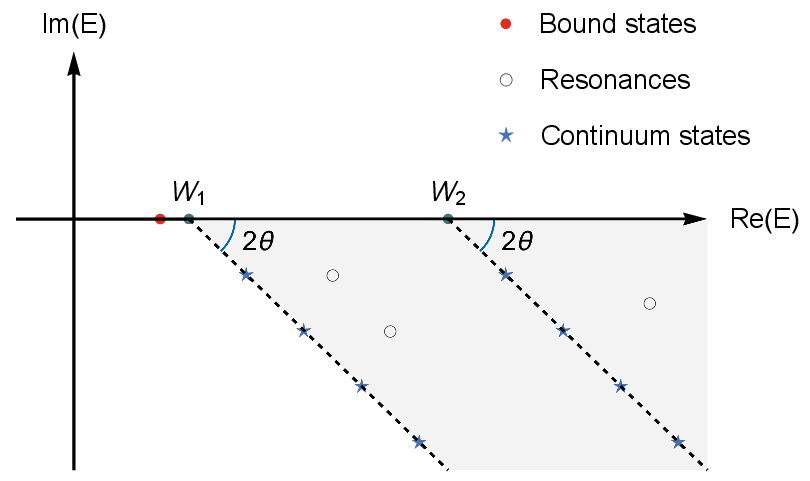}
    \caption{Schematic eigenvalue distribution of \( H^{\theta} \) in the coupled-channel two-body system. Continuum states are discretized  along the \( 2\theta \)-line as solid circles.
}
    \label{csmp}
\end{figure}

\subsection{Numerical solution methods}
In this work,  we use the Gaussian expansion method (GEM) to solve the coupled-channel Schr\"odinger equation, which is widely applied to solve the problem of a few-body system~\cite{Hiyama:2003cu,Hiyama:2018ivm}. In the GEM, the orbital wave functions are expanded in terms of a set of Gaussian basis functions
\begin{equation}{\label{wave1}}\phi_{n}(\boldsymbol{r}) =N_{nl_{r}}r^{l_{r}} e^{-\nu_{n}r^{2}} Y_{l_{r} m_{l_{r}}}\left(\boldsymbol{\hat{r}}\right),
 \end{equation} 
 \begin{equation}
\nu_{n}=1/r_{n}^{2}, r_{n}=r_{1}a^{n-1}(n=1,...,n_{max}),
\end{equation}
where the Gaussian parameters $\{r_1,~r_{max},~n_{max}\}$ = $\{0.5~\rm fm,~25~\rm fm,~25\}$ are applied in the present work.
In order to obtain resonance, the complex scaling method (CSM) is introduced.  In the CSM, the Hamiltonian $\hat{H}$ and wave function  $\mid\Phi\rangle$ are transformed with  complex rotation operator $\hat{U}(\theta)$
\begin{equation}
    \hat{H}_{\theta}=\hat{U}(\theta)\hat{H}\hat{U}^{-1}(\theta),\mid\Phi_{\theta}\rangle=\hat{U}(\theta)\mid\Phi \rangle.
\end{equation}
The coordinate $\bm{r}$ and conjugate momentum $\bm{p}$ are transformed to be
\begin{equation}
    \bm{r}\to\bm{r}e^{i\theta},\bm{p}\to\bm{p}e^{-i\theta},
\end{equation}
where the $\theta$ is variable and called scaling angle.
Correspondingly, the complex scaling Schr\"odinger equation can be wirtten as 
\begin{widetext}\begin{equation*}\begin{aligned}
    \left(\begin{array}{cccc}\label{seq}
       e^{-2i\theta}T_{11}+V_{11}(re^{i\theta})+W_{1} &V_{12}(re^{i\theta})&\cdots&V_{1j}(re^{i\theta})  \\
       V_{21}(re^{i\theta}) &e^{-2i\theta}T_{22}+V_{22}(re^{i\theta})+W_{2}&\cdots&V_{2j}(re^{i\theta})\\ 
      \vdots&\vdots&\vdots&\vdots\\ 
    V_{j1}(re^{i\theta})  & V_{j2}(re^{i\theta})&\cdots&e^{-2i\theta}T_{jj}+V_{jj}(re^{i\theta})+W_{j}\\
    \end{array}\right)\cdot\left(\begin{array}{c}
       \psi_{1}(r)   \\
        \psi_{2}(r)\\
        \vdots\\
        \psi_{j}(r)\\
    \end{array}\right) =E\left(\begin{array}{c}
       \psi_{1}(r)   \\
        \psi_{2}(r)\\
        \vdots\\
        \psi_{j}(r)\\
    \end{array}\right).
    \end{aligned}
 \end{equation*}
 \end{widetext}
The operator $T_{ij}$ s defined as
\begin{eqnarray}
    T_{ij}=\frac{1}{2\mu_{i}}\left(-\frac{d^{2}}{dr^{2}}+\frac{l_{i}(l_{i}+1)}{r^{2}}\right),
\end{eqnarray}
where $W_{ij}$	
  represents the corresponding threshold. As plotted in Figure~\ref{csmp}, on the complex-energy plane, the eigenvalues of the scattering continuum states align along the so-called $2\theta$ line, which satisfies the relation $1/2\text{Arg}(\Gamma/2\text{E})$. In contrast, bound states and resonances remain independent of the scaling angle $\theta$. This property allows one to distinguish bound states and resonances from the continuum states.

For a resonance, the RMS radius $r_{RMS}$ and the composed proportion $P$ are given by the c-product~\cite{T.yo,Lin:2023ihj}
\begin{eqnarray}
	r_{RMS}^2=(\psi^{\theta}|r^2|\psi^{\theta})=\sum_i\int r^2\psi^{\theta}_i(\bm{r})^2d^3\bm{r},\nonumber\\
	P=(\psi^{\theta}_i|\psi^{\theta}_i)=\int \psi^{\theta}_i(\bm{r})^2d^3\bm{r}\label{eq:c-prod},
\end{eqnarray}
where the $\psi_i$ is the wave function of the $i$-th channel and obeys
the normalization condition
\begin{eqnarray}
	\sum_i(\psi^{\theta}_i|\psi^{\theta}_i)=1.
\end{eqnarray}
The  $r_{\rm RMS}$ of resonant states can be a complex number. In such instances, one may apply the framework introduced by T. Berggren, which extends the concept of expectation values from bound states to resonances~\cite{Berggren:1970wto}. According to this framework, the real part of the complex $r_{\rm RMS}$ corresponds to the standard expectation value, while the imaginary part reflects the uncertainty associated with the measurement. This generalized interpretation has been supported by numerical calculations of $r^2$~\cite{Gyarmati:1972yac, 1997matrix}. Additionally, the $P$ is related to the square of the complex-scaled wave function $\psi^{\theta}_i(\bm{r})$, rather than its modulus squared. As a result, $P$ is not necessarily positive definite. Depending on the nodal structure and phase of the wave function, $P$ may take on positive or negative real values, or even possess an imaginary component.

\section{Results and discussions}\label{results}

\begin{figure}
    \centering
    \includegraphics[width=1\linewidth]{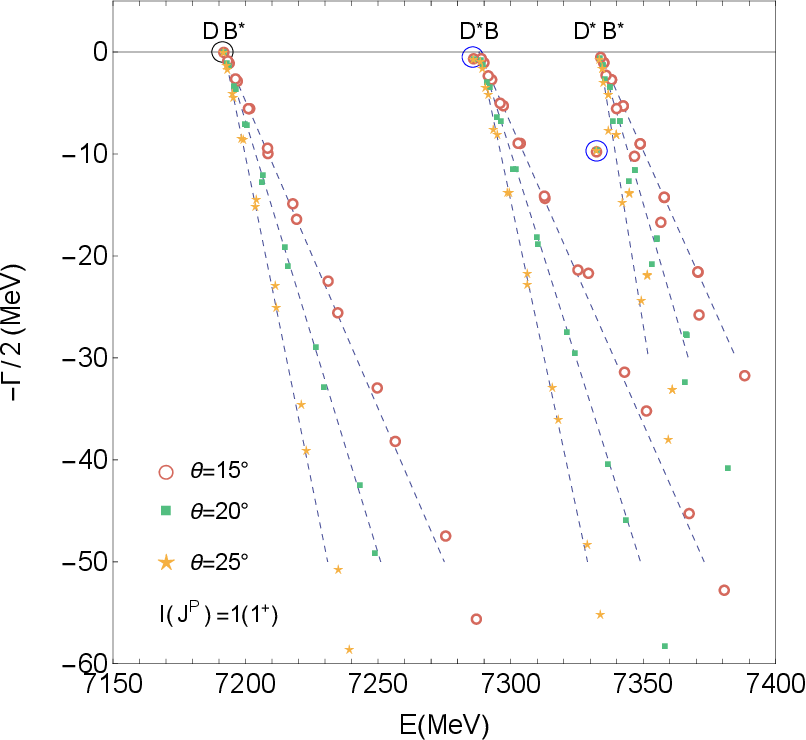}
    \caption{The complex energy eigenvalues of $I(J^P)=1(1^+)$ system with varying the angle $\theta$ from $15^\circ\sim25^\circ$.}
    \label{angle}
\end{figure}
\begin{figure*}
    \centering
\includegraphics[width=1\linewidth]{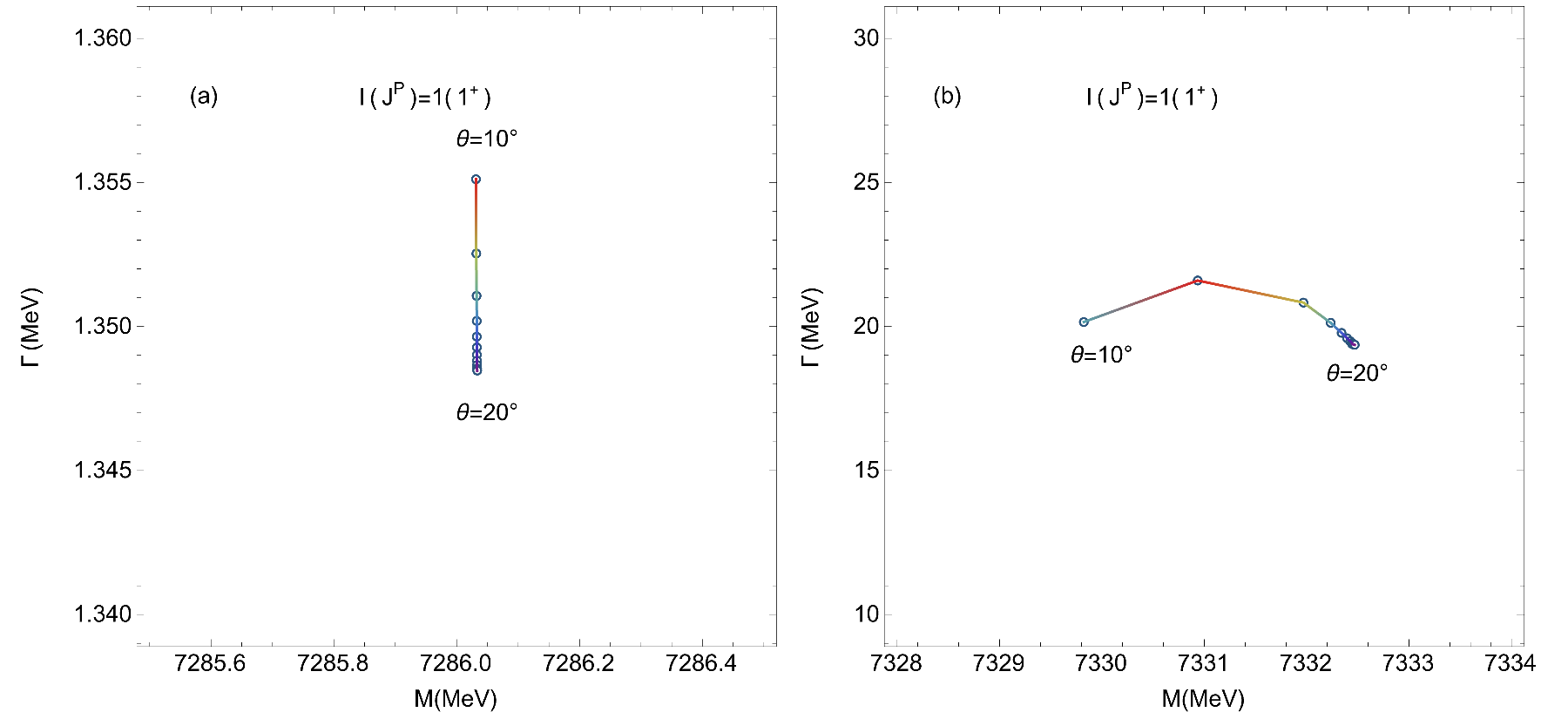}
    \caption{The variations of complex energies for the two predicted $1(1)^{+}$ resonances with $\theta$.}
    \label{circc}
\end{figure*}
\begin{figure*}
	\centering
	\includegraphics[scale=0.63]{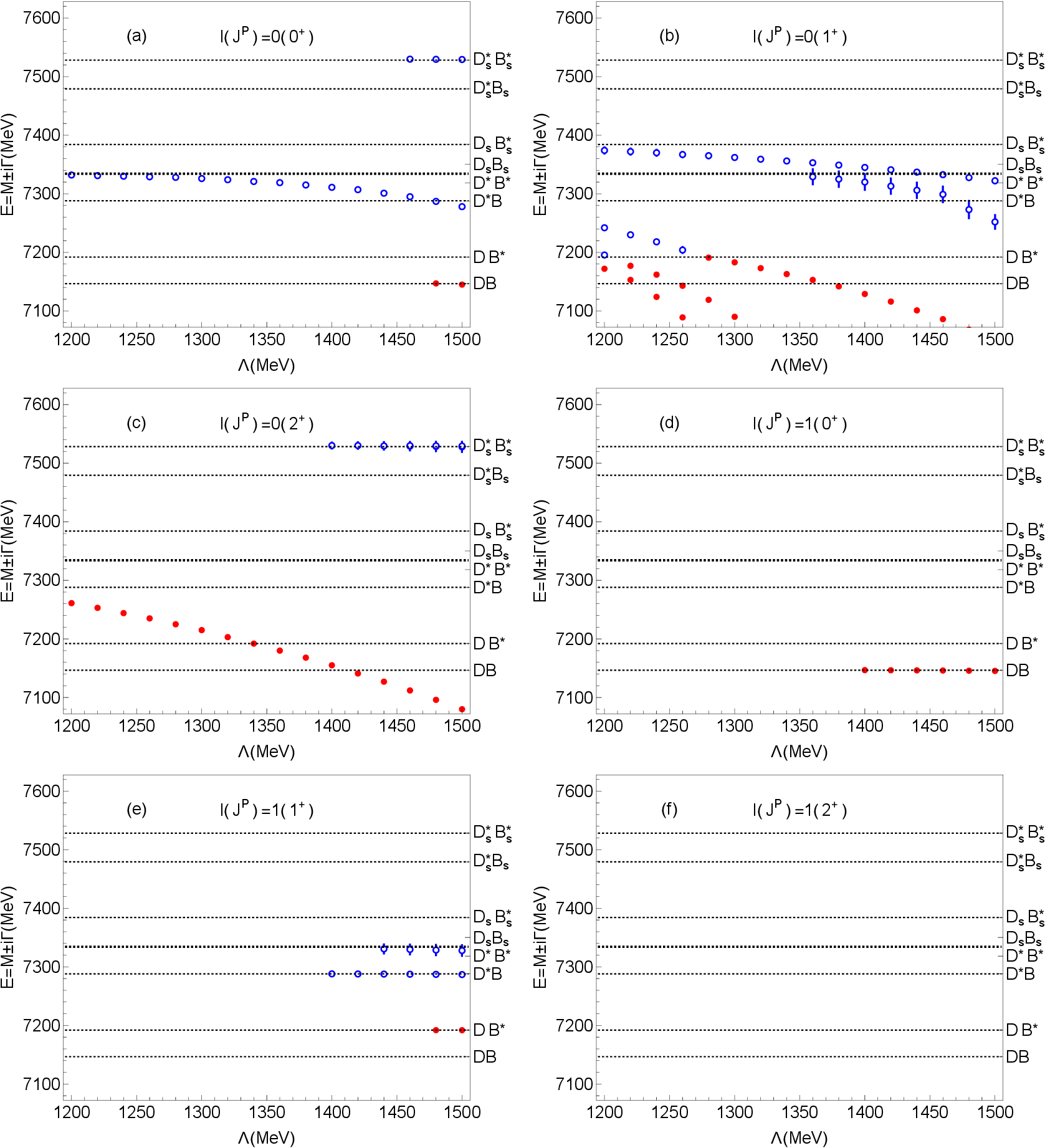}
	\caption{The $\Lambda$  dependence for the $D^{(*)}_{(s)}B^{(*)}_{(s)}$ systems. The blue open circles with bars correspond to the resonances, and the bars represent the total widths for resonances.}
	\label{bc}
\end{figure*}

\subsection{Isospin vector systems}

\begin{table*}[!htbp]
	\renewcommand\arraystretch{1.4}
	\caption{\label{zb} The molecular states for $I(J^{P})=1(1^{+}) $ system. The numbers in the bracket are the components proportion of $ DB^\ast+h.c.$, $ D^\ast B+h.c.$,  and  $D^\ast B^\ast $channels.}
	\begin{ruledtabular}
		\begin{tabular}{cccccccccccc}
			&$\Lambda(\rm{MeV})$
			&$E(\rm{MeV})$&$r_{RMS}(\rm{fm})$
			&($DB^{*}(^{3}S_{1})$&$DB^{*}(^{3}D_{1})$&$D^{*}B(^{3}S_{1})$&$D^{*}B(^{3}D_{1})$&$D^{*}B^{*}(^{3}S_{1})$&$D^{*}B^{*}(^{3}D_{1})$&$D^{*}B^{*}(^{5}D_{1})$)
			\\\hline
			&1400&$7287.71-0.36i$&$3.92-1.23i$& ($0.00-0.05i$&$-0.19+0.25i$&$98.38-0.49i$&$0.00+0.01i$&$1.68+0.24i$&$0.03+0.00i$&$0.10+0.04i$)\\\hline
			&\multirow{2}{*}{1450}&$7287.07-0.53i$&$2.74-1.11i$& ($0.00-0.07i$&$-0.23+0.47i$&$96.71-0.70i$&$0.01+0.00i$&$3.30+0.24i$&$0.04+0.01i$&$0.17+0.05i$)\\
			&&$7333.22-7.46i$&$1.22 -0.21i$& ($-0.80-1.35i$&$-0.33-0.13i$&$-1.86-2.66i$&$-0.13+0.49i$&$102.16+3.96i$&$0.93-0.26i$&$0.03-0.05i$)\\\hline
		&\multirow{3}{*}{1500}
			&$7286.03-0.67i$& $2.11 -0.75i$&($0.00-0.07i$&$-0.19+0.71i$&$94.53-0.74i$&$0.02+0.00i$&$5.36+0.03i$&$0.04+0.01i$&$0.24+0.04i$)\\
            &&$7332.50-9.68i$&$1.23 -0.22i$&($-1.11-1.64i$&$-0.39-0.12i$&$-4.53-2.32i$&$-0.08+0.57i$&$105.02+3.99i$&$1.06-0.42i$&$0.03-0.06i$)\\
            &&$7191.81$& $3.86$&(99.54&0.00&0.10&0.07&0.22&0.01&0.06)\\
		\end{tabular}
	\end{ruledtabular}
\end{table*}
By following the above procedures, one can explore the possible bound states and resonances of the $D_{(s)}^{(*)}B_{(s)}^{(*)}$ systems with various combinations by solving the coupled-channel Schr\"odinger equation using the complex scaling method. In this work, the only free parameter is the UV cutoff $\Lambda$ in Eq.~\eqref{FF}, which reflects the internal structure of the interacting hadrons. This parameter may vary for different coupled systems and lies within the range of $800 \sim 5000$ MeV. A reasonable cutoff for deuteron studies is estimated to be around 1000 $\rm MeV$, but it may vary across different scenarios~\cite{Tornqvist:1993ng,Tornqvist:1993vu}. It is worth noting that in Ref.~\cite{Chen:2022dad}, the authors studied the $D_{(s)}^{(*)} \bar{D}_{(s)}^{(*)}$ interactions, revealing charmonium-like resonances with quantum numbers $I(J^{PC})=0(0^{++})$ and $0(1^{+-})$ for cutoff values between $1000 \sim 2000$ $\rm~MeV$, with poles observed around 1500 $\rm~MeV$. In our previous work~\cite{Song:2024ngu}, we predicted several exotic hadron states for the $B_{(s)}^{(*)}\bar{B}_{(s)}^{(*)}$ systems by varying the cutoff in the range $800 \sim 1200$ $\rm{MeV}$. Since the interaction potential between the mesons is similar, the heavier mesons effectively reduce the contribution from the kinetic energy. Consequently, the cutoff value for the \( D^{(*)}_{(s)}B^{(*)}_{(s)} \) systems is expected to lie between that of the \( B_{(s)}^{(*)}\bar{B}_{(s)}^{(*)} \) systems and the \( D_{(s)}^{(*)}\bar{D}_{(s)}^{(*)} \) systems. According to heavy quark symmetry, exotic hadron states should exist that correspond to the partners of the charmonium-like and bottomonium-like states in the bottom-charmed sector. Therefore, we first investigate the $I(J^{P})=1(1^+)$ $DB^*/D^*B/D^*B^*$ channel, which helps determine the cutoff value within a phenomenologically reasonable range.

For the \( I(J^{P})=1(1^+) \) \( DB^*/D^*B/D^*B^* \) channel, no poles are found when the cutoff \(\Lambda\) varies from 800 to 1300~$\rm MeV$. However, at \(\Lambda = 1400~\rm MeV\), a narrow resonance appears near the \( D^*B \) threshold, predominantly composed of the \( D^*B(^3S_1) \) component. When the cutoff increases to 1450~$\rm MeV$, another resonance emerges below the \( D^*B^* \) threshold with a complex energy of \( E - i\Gamma/2 = 7333 - 7i~\rm MeV \). This resonance is mainly dominated by the \( D^*B^*(^3S_1) \) channel, and may be considered a promising partner to both the \( T_{c\bar{c}1}(4020) \) and \( T_{b\bar{b}1}(10650) \) states. As the cutoff further increases to around 1500~$\rm MeV$, a loosely bound state with a predicted mass of 7192~$\rm MeV$ is found, which could correspond to the \( T_{c\bar{c}1}(3900) \) and \( T_{b\bar{b}1}(10610) \) states. It is noteworthy that, for similar cutoff values, previous studies considering only single-channel interactions identified solely bound states~\cite{Sun:2012sy}, highlighting the crucial influence of coupled-channel effects on molecular state formation. 

In Figure~\ref{angle}, we present the complex energy eigenvalues for the \( I(J^P)=1(1^+) \) system at \( \Lambda = 1500~\mathrm{MeV} \), with the scaling angle \( \theta \) varied from \( 15^\circ \) to \( 25^\circ \). As shown, the bound states (enclosed by black circles) and resonances (enclosed by blue circles) remain invariant with respect to the scaling angle \( \theta \), whereas the continuum states rotate along the so-called \( 2\theta \) line. In Figure~\ref{circc}, we illustrate the dependence of the complex energies of the two predicted resonances on \( \theta \), showing that numerical errors induced by the choice of \( \theta \) are minimal and quickly converge as \( \theta \) increases. The detailed numerical results are summarized in Table~\ref{zb} and Figure~\ref{bc}(e).

For the three predicted exotic states, the root-mean-square radius (\( r_{\mathrm{RMS}} \)) is consistent with that of typical hadronic molecules such as the deuteron. Therefore, a cutoff range of \( 1200 \sim 1500~\rm MeV \) is adopted for investigations of other \( D_{(s)}^{(*)}B_{(s)}^{(*)} \) systems. The corresponding numerical results, depending on the cutoff parameter \(\Lambda\), are presented in Figure~\ref{bc}.

For the \( I(J^{P})=1(0^{+}) \) \( DB/D^{*}B^{*} \) systems, a loosely bound state is predicted when the cutoff \(\Lambda\) is within the range of \( 1400 \sim 1500 \)$\rm~MeV$. In this region, the root-mean-square radius $r_{RMS}$ decreases from 4 to 2 $\rm fm$, suggesting that it can be considered a promising molecular state candidate. However, for the \( D^{*}B^{*} \) system with spin-parity \( I(J^P) = 1(2^+) \), no structure is observed in the complex energy plane when the cutoff is set to \( 1200 \sim 1500 \)$\rm~MeV$. It is important to note that the \( I(J^P) = 1(2^+) \) system consists solely of the \( D^{*}B^{*} \) channel, where the attractive interaction is not strong enough to form a bound state compared to other isospin-vector systems. Nevertheless, by increasing the cutoff to approximately 2500$\rm~MeV$, one may explore the possibility of a bound state.  

\subsection{Isospin scalar systems}

Except for isospin vector systems, we also have isospin scalar systems for $D^{(*)}_{(s)}B^{(*)}_{(s)}$. Unlike the \( I=1 \) systems, the \( I=0 \) case allows for more coupled channels. As shown in Figure~\ref{bc}(a), we investigate the coupled-channel system \( DB/D^*B^*/D_{s}B_{s}/D_{s}^{*}B_{s}^{*} \) with spin-parity \( I(J^P)=0(0^+) \). A narrow resonance, predominantly composed of \( D^*B^*(^1S_{0}) \)($58.78 -7.24i$) and \( D_sB_s(^1S_{0}) \)($29.74 +6.81i$), appears near the \( D_sB_s \) threshold with a complex energy of \( E-i\Gamma/2=7332-0.5i \)$\rm~MeV$ when the cutoff value is set to \( \Lambda = 1200 \)$\rm~MeV$. By increasing the cutoff, a bound state solution emerges below the \( DB \) threshold, meanwhile, the resonance evolves to \( 7387-0.3i \)$\rm~MeV$. Both states can be considered  candidates for exotic hadronic states. Additionally, we identify a pole near the \( D_{s}^{*}B_{s}^{*} \) threshold with a narrow width.  

For the \( I(J^P) = 0(1^+) \) channel, the dependence of the system on the cutoff parameter is illustrated in Figure~\ref{bc}(b). As the cutoff increases from \( 1200 \) to \( 1500~\mathrm{MeV} \), several notable features are observed. When the cutoff lies within the range \( 1200 \sim 1300~\mathrm{MeV} \), two poles are excluded, as their corresponding \( r_{\text{RMS}} \) values do not support a molecular interpretation. In particular, a cutoff-sensitive resonance located below the \( D^*B \) threshold evolves into a bound state as the cutoff increases to \( 1280~\mathrm{MeV} \); however, due to its relatively small \( r_{\text{RMS}} \) in the range \( 1300 \sim 1500~\mathrm{MeV} \), it is not regarded as a suitable molecular candidate in the present analysis. Furthermore, when the cutoff increases to \( 1320~\mathrm{MeV} \), a new resonance with energy \( 7323 - 4i~\mathrm{MeV} \), predominantly composed of the \( D^*B^*(^3S_1) \) component, emerges and can be interpreted as a possible partner of the \( X(4014) \) particle. Meanwhile, for the pole located below the \( D_sB_s^* \) threshold, its complex energy \( E - i\Gamma/2 \) shifts from \( 7374 - 5i~\mathrm{MeV} \) to \( 7322 - 4i~\mathrm{MeV} \) as the cutoff increases from \( 1200 \) to \( 1500~\mathrm{MeV} \), indicating a gradual stabilization of this state.

Finally, for the \( I(J^P) = 0(2^+) \) \( D^*B^*/D_s^*B_s^* \) system, a deeply bound state is found at \( \Lambda = 1200~\mathrm{MeV} \). However, its corresponding \( r_{\text{RMS}} \) value is too small to support a molecular interpretation. As the cutoff increases to approximately \( \Lambda = 1400~\mathrm{MeV} \), a resonance with a predicted energy of \( E - i\Gamma/2 = 7530 - 5i~\mathrm{MeV} \) emerges near the \( D_s^*B_s^* \) threshold. This state is predominantly composed of the \( D_s^*B_s^*(^3S_2) \) component and, according to our analysis, can be considered a promising molecular candidate.

   \begin{table*}[!htbp]
 	\renewcommand\arraystretch{1.4}
 	\caption{\label{sum} The summary of our predictions for $D^{(*)}_{(s)}B^{(*)}_{(s)}$ systems with cutoff $\Lambda$ in a range of $1200\sim1500$ $\rm~MeV$. Here, the $"-"$ means nonexistence, $"\checkmark"("\times")$ represents that the corresponding state may (may not) form a molecular state.}
 	\begin{ruledtabular}
 		\begin{tabular}{ccccccccc}
 			&$I(J^{PC})$
 			&Mass$(\rm$\rm~MeV$)$&Width$(\rm$\rm~MeV$)$&$r_{RMS}(\rm fm)$&Status&Selected decay mode
 			\\\hline
 			&\multirow{3}{*}{$0(0^{+})$}&$7326.00\sim7277.68$&$0.76\sim0.82$&$3.94 -1.26 i\sim1.23 -0.30 i$&$\checkmark$&$B_c\eta^{(\prime)}$/$B_c^*\omega$/$DB$\\
            		&&$7529.63\sim7528.91 $&$6.16\sim6.48$&$4.64 +0.73i\sim3.52+1.04 i$&$\checkmark$&$B_c\eta^{(\prime)}$/$B_c^*\phi$/$D_s^{(*)}B_s^{(*)}$\\
               		&&$7146.74\sim7145.49$&$-$&$3.53\sim2.60$   &$\checkmark$&$B_c\eta^{(\prime)}$/$B_c^*\omega$ \\              \hline
 			&\multirow{3}{*}{$0(1^{+})$}&$7330.73 \sim7251.56 $&$24.06\sim23.62$&$4.74 -1.07 i\sim2.89 -0.37 i$&$\checkmark$&$B_c\omega$/$B_c^*\eta^{(\prime)}$/$DB^*$/$D^*B$ \\
 			&&$7374.1 0\sim7322.14$&$10.48\sim8.48$&$2.32 +0.08i\sim2.40 +0.62i$&$\checkmark$&$B_c\omega$/$B_c^*\eta^{(\prime)}$/$DB^*$/$D^*B$\\&&$7182.86\sim7049.72$&$-$&$1.00\sim0.48$&$\times$ &$-$\\\hline	
 			&\multirow{2}{*}{$0(2^{+})$}&$7214.52\sim7079.97$&$-$&$0.52\sim0.42$&$\times$&$-$\\
            		&&$7530.45\sim7527.80$&$9.46\sim17.24$&$4.20+1.51i\sim1.15 -0.86 i$&$\checkmark$&$B_c\phi$/$B_c^*\phi$/$D_s^{(*)}B_s^{(*)}$  \\\hline
&$1(0^{+})$&$7146.69\sim7145.31$&$-$&$4.33\sim2.46$&$\checkmark$&$B_c\pi$/$B_c^*\rho$\\\hline
 			&\multirow{3}{*}{$1(1^{+})$}&$7288.05\sim7286.03$&$0.58\sim1.34$&$3.92- 1.23i\sim2.11 - 0.75i$&$\checkmark $  &$B_c\rho$/$B_c^*\pi$/$DB^*$  \\&&$7333.25\sim7332.39$&$14.92\sim19.6$&$1.22 -0.21i\sim1.23 +0.26 i$&$\checkmark$&$B_c\rho$/$B_c^*\pi$/$DB^*$/$D^*B$   \\
            &&$7191.94\sim7191.83$&$-$&$4.47\sim4.19$&$\checkmark$&$B_c\rho$/$B_c^*\pi$  \\\hline
 			&$1(2^{+})$&$-$&$-$&$-$&$\times$&$-$ \\
 		\end{tabular}
 	\end{ruledtabular}
 \end{table*}

\subsection{Further discussion}
According to our calculations, owing to the inclusion of more allowed channels in the present work, both bound states and resonances are obtained for the \( I=0 \) and \( I=1 \) \( D^{(*)}_{(s)}B^{(*)}_{(s)} \) systems. In contrast, only bound states were reported in the hidden-bottom molecular tetraquark system with \( I=1 \) in Ref.~\cite{Song:2024ngu}. This highlights the crucial role of coupled-channel effects in the formation of molecular states. For instance, when considering the single-channel \( DB^* \) or \( D^*B \) system with \( I(J^P) = 1(1^+) \), no pole is found for cutoffs ranging from \( 800 \) to \( 1500~\mathrm{MeV} \), as the interaction is too weak to generate a molecular state. However, if only the \( I(J^P)=1(1^+) \) \( D^*B^* \) channel is considered, a bound state appears at a cutoff around \( 1300~\mathrm{MeV} \), consistent with the result of Ref.~\cite{Sun:2011uh}. Moreover, even when the contributions from the \( D \)-wave are neglected, bound and resonant states still emerge at sufficiently large cutoff values. This further demonstrates that the coupled-channel effect is dominant in the formation of resonant states, with the main contributions originating from the \( S \)-wave interactions.

Moreover, we prefer to collect the final results of $D^{(*)}_{(s)}B^{(*)}_{(s)}$ systems in Table~\ref{sum}, where we predict ten exotic particles. As plotted in Figure~\ref{bc}(b) and Table~\ref{sum}, for $I(J^P)=0(1^+)$ system, we predict two exotic states near $D^*B$, and $D^*B^*$ thresholds, respectively, which may be regarded as good molecular state and be counterparts to the exotic states already observed in both the charm and bottom sectors. Also, for $I(J^P)=1(1^+)$ system, three molecular states are found below their thresholds, respectively, which are shown in figure~\ref{bc}(e), Table~\ref{zb}, and Table~\ref{sum}. However, whether the $T_{b\bar{b}1}(10610)$ and $T_{b\bar{b}1}(10650)$ are located above or below the corresponding thresholds is still an open question. For the bottom-charmed molecular tetraquark states, they may also encounter the same question, thus, the further theoretical studies and experimental explorations are needed. Finally, according to the masses
and quantum numbers, we collect some possible decay modes
of these predicted particles in Table~\ref{sum}. It can be evident that the
$B_c/B^*_c$ plus light mesons are the ideal final states
to search for the bound states, while the $D_{(s)}^{(*)}B_{(s)}^{(*)}$
channels are suitable for the resonances. We highly recommend that the future experiments could hunt for these bottom-charmed exotic particles in the future.

\section{Summary}\label{sec4}
 
In this work, we investigate the \( D^{(*)}_{(s)}B^{(*)}_{(s)} \) systems from a coupled-channel perspective within the one-boson-exchange (OBE) model. By employing the Gaussian expansion method and the complex scaling method, we solve the coupled-channel Schr\"odinger equation and identify ten exotic hadronic states. We first analyze the \( D^{(*)}_{(s)}B^{(*)}_{(s)} \) system with quantum numbers \( I(J^P) = 1(1^+) \), where several poles are found and denoted as \( T_{c\bar{b}1} \), which may be interpreted as partners of the previously observed exotic states \( T_{c\bar{c}1} \) and \( T_{b\bar{b}1} \). Subsequently, we explore various \( D^{(*)}_{(s)}B^{(*)}_{(s)} \) configurations with different quantum numbers to search for additional bound and resonant states. In particular, one molecular candidate is predicted near the \( D^*B^* \) threshold in the \( I(J^P) = 0(1^+) \) sector. We expect that our results provide useful theoretical guidance and strongly encourage future experimental searches for these exotic hadronic molecules.

\subsection*{ACKNOWLEDGMENTS}

We would like to thank Dian-Yong Chen for useful discussions. The work of X.-N. X. and Q. F. Song is supported by the National Natural Science
Foundation of China under Grants No. 12275364.  Q.-F. Lü is supported by the Natural Science Foundation of Hunan Province under Grant No. 2023JJ40421, the Scientific Research Foundation of Hunan Provincial Education Department under Grant No. 24B0063, and the Youth Talent Support Program of Hunan Normal University under Grant No. 2024QNTJ14. Wei Liang is supported by the Fundamental Research Funds for the Central Universities of Central South University under Grants No. 1053320214315 and No. 2022ZZTS0169, the Postgraduate Scientific Research Innovation Project of Hunan Province under Grant No. CX20220255.

\end{document}